\documentclass[aps]{revtex4}
\usepackage{amsmath}
\usepackage{amssymb}
\usepackage{epsfig}
\usepackage{amscd}
\usepackage{graphicx,xspace}
\usepackage{float}
\newcommand{\mpi}{\pi}
\newcommand{\mi}{\mathrm{i}}
\newcommand{\period}{\text{per}}
\newcommand{\flabel}[1]{\label{f:#1}}
\newcommand{\elabel}[1]{\label{e:#1}}
\newcommand{\slabel}[1]{\label{s:#1}}
\newcommand{\eq}[1]{eqn~(\ref{e:#1})}
\newcommand{\eqq}[1]{Equation~(\ref{e:#1})}
\newcommand{\eqtwo}[2]{eqns~(\ref{e:#1}) and~(\ref{e:#2})}
\newcommand{\fig}[1]{Fig.~\ref{f:#1}}
 
\newcommand{\sect}[1]{Section~\ref{s:#1}} 
\newcommand{\app}[1]{Appendix~\ref{s:#1}} 
\newcommand{\GAUSS}{{\text{Gauss}}}  
\newcommand{\Tr}{\text{Tr}}  

\newcommand{\PCAL}{\mathcal{P}}  
\newcommand{\SCAL}{\mathcal{S}}  
\newcommand{\SET}[1]{\{#1\}}
\newcommand{\expl}[1]{\exp \lb #1 \rb } 
\newcommand{\expa}[1]{\mathrm{e}^{#1}}   
\newcommand{\expb}[1]{\exp \glb #1 \grb} 
\newcommand{\expc}[1]{\exp \glc #1 \grc} 
\newcommand{\expd}[1]{\exp \gld #1 \grd} 
\newcommand{\sina}[2][]{\sin^{#1} \! \gla #2 \gra}  

\newcommand{\sinb}[2][]{\sin^{#1} \glb #2 \grb}  
\newcommand{\cosb}[2][]{\cos^{#1} \glb #2 \grb}  
\newcommand{\cotb}[2][]{\cot^{#1} \glb #2 \grb}  

\newcommand{\cothc}[2][]{\coth^{#1} \glc #2 \grc} 

\newcommand{\lb}{\left[}  
\newcommand{\rb}{\right]}  

\newcommand{\gla}{\,}  
\newcommand{\gra}{}  
\newcommand{\glb}{\left(}  
\newcommand{\grb}{\right)}  
\newcommand{\glc}{\left[}  
\newcommand{\grc}{\right]}  
\newcommand{\gld}{\left\{}  
\newcommand{\grd}{\right\}}  

\newcommand{\const}{\text{const}}
\newcommand{\TO}{,\ldots,}
\newcommand{\dd}[1]{\text{d}{#1\ }}   
\newcommand{\ddd}[1]{\text{d}{#1}}   

\newcommand{\mean}[1]{\left\langle #1 \right\rangle}
\newcommand{\half}{\frac{1}{2}}

\begin{document}
\title{Universal width distributions in non-Markovian Gaussian processes} 
\author{Raoul
Santachiara} \email{raoul.santachiara@lpt.ens.fr}
\affiliation{CNRS-Laboratoire de Physique Th\'eorique, Universit\'e
Louis Pasteur, \\ 3 rue de l'Universit\'e, 67084 Strasbourg Cedex,
France} 
\affiliation{CNRS-Laboratoire de Physique Th\'eorique,
Ecole Normale
Sup{\'{e}}rieure\\  24 rue Lhomond, 75231 Paris Cedex 05, France}
\author{Alberto Rosso} \email{rosso@lptms.u-psud.fr}
\affiliation{CNRS-Laboratoire de Physique Th\'{e}orique et Mod\`{e}les
Statistiques \\ B\^{a}t. 100 Universit\'{e} Paris-Sud; 91405 Orsay
Cedex, France} \author{Werner Krauth} \email{krauth@lps.ens.fr}
\affiliation{CNRS-Laboratoire de Physique Statistique \\ 
Ecole Normale
Sup{\'{e}}rieure; 24 rue Lhomond, 75231 Paris Cedex 05, France
}
\date{13 Nov 2006}

\begin{abstract}
We study the influence of boundary conditions on self-affine random
functions $u(t)$  in the interval $t/L \in [0,1]$, with independent
Gaussian Fourier modes of variance $\sim 1/q^{\alpha}$. We consider
the probability distribution of the mean square width of $u(t)$
taken over the whole interval or in a window $t/L \in [x, x+\delta]$.
Its characteristic function can be expressed in terms of the spectrum
of an infinite matrix.  This distribution strongly depends on the
boundary conditions of $u(t)$ for finite $\delta$, but we show that 
it is universal
(independent of boundary conditions) in the small-window limit ($\delta
\to 0$, $\delta \ll \min[x,1-x]$).  We compute it directly for all values
of $\alpha$, using, for $\alpha<3$, an asymptotic expansion formula that
we derive.  For $\alpha > 3$, the limiting width distribution
is independent of $\alpha$. It corresponds to an infinite matrix with a
single non-zero eigenvalue.  We give the exact expression for the width
distribution in this case.  Our analysis facilitates the estimation
of the roughness exponent from experimental data, in cases where the
standard extrapolation method cannot be used.
\end{abstract}
\maketitle
\section{Introduction}

In nature, random processes model interfaces and surfaces
\cite{toroczkai} \cite{kardar}, turbulent flows \cite{frisch}, erratic
time series\cite{surya} and many other systems.  In the most simple
setting, these random processes correspond to a function of a scalar
variable $u(t)$, and are characterized by a probability distribution
$\PCAL[u(t)]$ that stems from an equilibrium problem or a
non-equilibrium process. The probability distribution $\PCAL[u(t)]$ is
often unknown, and generally inaccessible to exact analysis. In many
cases, approximate methods must be brought to bear on these problems.

One of the most successful approaches in the field of random processes
is the Gaussian approximation. It consists, schematically, in
developing $u(t)$ on a general Fourier basis,
\begin{equation}
   u(t) \sim \sum_{n=-\infty}^{\infty} 
   a_n \expb{\mi q_n  t}, 
   \elabel{rough_equation}
\end{equation}
and in assuming that the coefficients $a_n= a^*_{-n}$ are independent
Gaussian random variables. The probability distribution
$\PCAL[u(t)]=\PCAL[\SET{a_1, a_2, \dots}]$ then factorizes into a
product $\PCAL \simeq \PCAL^\GAUSS = \PCAL(a_1)\PCAL(a_2)\dots$, where
$\PCAL(a_n)$ is a Gaussian with zero mean and variance
$\sigma^2_n$. The idea behind this representation is to approximate
the translation-invariant action of a complicated interacting problem
by a quadratic (Gaussian) action which yields the same two-point
correlation functions as in the original problem.  In the particularly
important case of self-affine (critical) systems, the only length
scale present is the system size.  The scaling of the variances with
the modes $q_n$ can then be described by a single parameter $\alpha$
with $\sigma^2_n \propto 1/n^{\alpha}$. In many non-trivial problems,
the Gaussian approximation is in outstanding quantitative agreement
with the full theory \cite{rosso_width,ledoussal_Rcalcul}, and reproduces
very well even the higher-order correlation functions.

Because of its decoupled Fourier modes, the Gaussian approximation
is considerably simpler than the full theory and the function $u(t)$
can easily be generated through \eq{rough_equation} from Gaussian
random numbers $\SET{a_n}$ (see \fig{samples}). However, the price
to pay for this simplicity in Fourier space is to have non-trivial
long-range correlations in real space. As we will discuss in detail,
the real-space action contains generalized derivatives which, in the
sense of the Riemann--Liouville derivative, can be expressed as an
integral convolution with a long-range kernel \cite{oldham_spanier}.
For non-integer values of $\alpha/2$, the real-space Gaussian action
is non-local and the geometrical properties of the function $u(t)$
are intricate.

For $\alpha=2$, the Gaussian approximation corresponds to the notorious
random walk. In this case the real-space action is local. It defines
a Markovian evolution (the value of $u$ at $t+\ddd{t}$ depends only on
the one at $t$). For this reason it is possible to determine in detail
its geometric properties (for a recent review see \cite{majumdar}).
For $\alpha\neq 2$, instead, the process $u(t)$ is non-Markovian: In the
case $\alpha = 4$ (the driven curvature model \cite{toroczkai,einstein}),
the evolution of the derivative $ \dot{u}(t)$ is Markovian, but not the
one of $u(t)$ itself. For not-integer $\alpha/2$, the non-Markovian
properties reflect the non-local character of the action.  This
implies that, generally, memory effects influence the shape of $u(t)$
and calculations, even within the Gaussian approximation,  are difficult. In
particular, the persistence exponents and the distribution of the extreme
remain unknown \cite{majumdar_bray,majumdar_alain}. On the other hand,
the boundary conditions influence the statistical properties of $u(t)$
in a non-trivial way for all $t \in [0,L]$.
Understanding these effects is important because the periodic
boundary conditions for $u(t)$, which are commonplace in theoretical
calculations and in numerical simulations, are usually not realized in
experiments \cite{dequeiroz,rosso_width2}.

\begin{figure}
   \centerline{
   \epsfxsize=12.0cm
   \epsfclipon
   \epsfbox{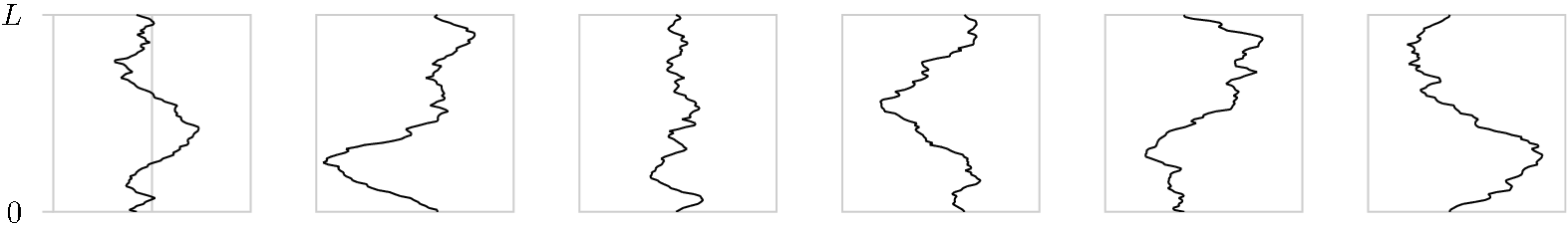}}
   \caption{Gaussian functions $u(t)$ in the interval $ t/L \in [0,1]$,
   corresponding to a probability distribution $\PCAL^\GAUSS[u(t)]$
   with $\zeta=0.75$ ($\alpha=2.5$) (full periodic series).}
   \flabel{samples}
\end{figure}
As an example of a fundamental geometrical quantity sensitive to the
boundary, we consider the mean square width
\begin{equation*}
w_2= \frac{1}{L} \int_0^{L}  \dd{t} u^2(tL) 
- \frac{1}{L^2}\glc \int_0^{L} \dd{t} u(tL)  \grc^2, 
\end{equation*}
which is relevant both from the theoretical and the experimental points
of view.  For a self-affine random process, the mean square width is
itself described by  a non-trivial probability distribution $\PCAL(w_2)$, 
which is neither a delta-function nor a Gaussian.

In this work, we study the influence of boundary conditions on the
width distribution of $\PCAL^\GAUSS[u(t)]$ for self-affine functions
characterized by variances scaling with the single parameter $\alpha$.
In fact, the schematic Fourier representation of \eq{rough_equation}
can be rendered explicit in a number of ways in order to accommodate the
boundary conditions. First it can be expanded into a full periodic
series,
\begin{equation}
   u(t)= \sum_{n=1}^{\infty} a_n \cosb{\frac{2 \mpi n}{L}t} + b_n
    \sinb{ \frac{2 \mpi n}{L}t }\ \text{(full periodic series)}, 
   \elabel{full_period_series}
\end{equation}
where $a_n$ and $b_n$ are independent Gaussian random numbers of
variance $1/n^\alpha$.  This implies that the function, which has zero
average value, and all the derivatives are periodic, if they exist ($
u^{(k)}(t=0)= u^{(k)}(t=L)$ for $k=0,1,\dots \le \alpha/2,$). It is
also possible to expand the Gaussian function $u(t)$ into a sine
Fourier series,
\begin{equation}
   u(t)= \sum_{n=1}^{\infty} s_n \sinb{ \frac{\mpi n}{L}t}\ \text{(sine series)},
   \elabel{sine_series}
\end{equation}
again supposing that the $s_n$ are independent Gaussians with variance
$\propto 1/n^\alpha$.  In that case, the function $u(t)$ vanishes at
$t=0$ and $t=L$. By a uniform shift of the function, it can be made
to have zero average value, as for the full periodic series. However,
all the existing even derivatives vanish ($u^{(k)}(0)= u^{(k)}(L)=0$
for $k=0,2,\ldots\le \alpha/2$) for the sine Fourier series. The full
periodic series and the sine series are equivalent for the random walk
($\alpha=2$) \cite{feymann,werner_book}, but they differ for all other
values of $\alpha$.

Finally, the function $u(t)$ can also be expanded into a cosine
Fourier series,
\begin{equation}
   u(t)= \sum_{n=1}^{\infty} c_n \cosb{\frac{\mpi n}{L}t}\
   \text{(cosine series)}, 
   \elabel{cosine_series}
\end{equation}
again with $\sigma^2(c_n) \propto 1/n^\alpha$.  In this case, the
random function is not forced to satisfy $u(0)=u(L)$, and for this
reason it has been used to study free random walks
\cite{majumdar}. Analogously to the sine Fourier series, all the odd
derivatives of $u(t)$ must vanish at the boundaries, if they exist
($u^{(k)}(0)= u^{(k)}(L)=0$ for $k=1,3,\ldots\le \alpha/2$). This
paper shows that these additional constraints strongly influence the
geometry of the function $u(t)$ for $\alpha \neq 2$.

The above three series representations correspond to the same real-space
action $\SCAL[u(t)]$ and its associated Gaussian probability distribution
$ \PCAL^\GAUSS \glc u(t) \grc$
\begin{equation}
   \SCAL[u(t)] = \frac{1}{2}\int_0^L \dd{t}
   \glb\frac{\partial^{\alpha/2}u(t)}{\partial t^{\alpha/2}} \grb^2
     \implies \PCAL^\GAUSS \glc u(t) \grc  = \expd{-\SCAL[u(t)]}.
   \elabel{action}
\end{equation}
For non-integer values of $\alpha/2$, the generalized
derivative in \eq{action} is defined in momentum space: $\glc
\partial^{\alpha/2}/\partial t^{\alpha/2} \grc \expa{\mi q t}=(\mi
q)^{\alpha/2} e^{\mi q t}$ (the derivative can also be defined  in real space \cite{oldham_spanier}). 
Using this definition we can compute the variances of the Fourier coefficients:  
\begin{equation}
\glc \sigma_n^{\period} \grc^2= \frac{L^{\alpha-1}}{2^{\alpha -1} \mpi^\alpha
  n^\alpha}  \implies
  \PCAL^\GAUSS[\{a_n,b_n\}] = 
  \prod_{n=1}^\infty \frac{1}{2 \mpi \sigma_n^2}
     \exp\glc - \frac{1}{2} \frac{(a_n^2+b_n^2)}{\sigma_n^2} \grc
   \quad \text{(full periodic series)}.
   \elabel{sigmafp}
\end{equation}
For the sine and the cosine series, the variances are
\begin{equation}
   \glc\sigma_n^{\cos} \grc ^2=\glc \sigma_n^{\sin} \grc^2= \frac{2L^{\alpha-1}}{ \mpi^\alpha n^\alpha}.
   \elabel{sigmasc}
\end{equation}
This choice leads to analogous expressions for the probability
distributions $ \PCAL^\GAUSS\glc \gld s_n \grd \grc $ and $\PCAL^\GAUSS \glc \gld  c_n \grd \grc  $, respectively.

The variances in \eqtwo{sigmafp}{sigmasc} scale with the system size
as $\propto L^{\alpha -1} = L^{2 \zeta}$ where $\zeta $ is the
roughness exponent.  This exponent characterizes the main geometric
properties of a self-affine system.

\begin{figure}[h]
   \centerline{
   \epsfxsize=10.0cm
   \epsfclipon
   \epsfbox{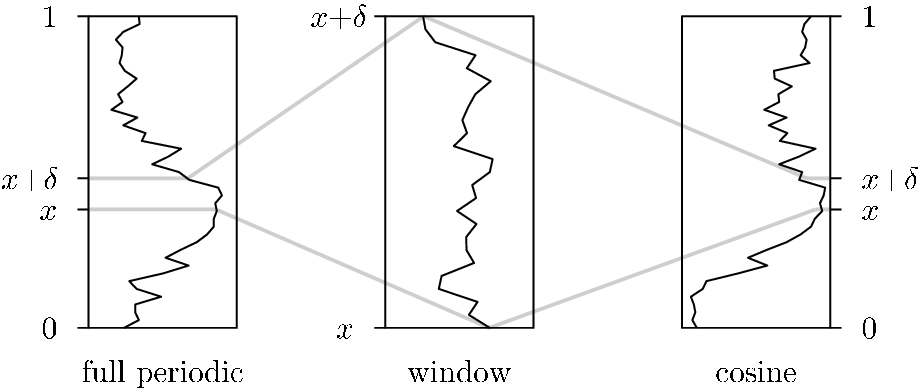}}
   \caption{Gaussian functions $u(t/L)$ corresponding to the full periodic and the cosine series.
       Window boundary conditions correspond to 
       picking out a piece of $u(t)$ in the interval $t/L \in [x,x+\delta]$ and
       shifting $u(t/L)$  such that the window mean value vanishes. }
   \flabel{three_models}
\end{figure}
In this paper, we first consider Gaussian functions on the entire interval $[0,L]$. 
We compute the
average of the width distribution for the three series
and for general values of $\alpha$ (\sect{avwidth1}). The average of the width
distribution strongly depends on the boundary conditions (for $\alpha \ne 2$).
These results are then generalized (\sect{avwidth2}) to the case when the
function $u(t)$ is restricted to a window of width $ \delta$, in the interval
$t/L \in [x, x + \delta]$:
\begin{equation}
   w_2(x,\delta)= \frac{1}{\delta}\int_x^{x+\delta} \dd{t} u^2(t L) 
   -\glb\frac{1}{\delta} \int_x^{x+\delta}\dd{t} u(t  L)  \grb^2.
   \elabel{w2_window}
\end{equation}
These window boundary conditions are closer to the experimental
situation than those realized by either the full periodic or the
cosine series (see \fig{three_models}).  We write (\sect{phi}) the characteristic function of
the width distribution in terms of the eigenvalues of an infinite matrix
which depends on the basis functions of the Fourier series, and on the
window parameters $\SET{\delta, x}$. Practically, to compute
the width distribution for any of the series, in an arbitrary finite window,
it suffices to compute this characteristic
function from the eigenvalues of a finite-rank approximation of the above
matrix, and to perform an inverse Fourier transform.  It is easy to see
that the width distribution  depends on the size of the window and on
the choice of boundary conditions.  In the small-window limit ($\delta
\rightarrow 0$ and far from the boundaries $\delta \ll \min[x,1-x]$),
finite-rank approximations to the above matrix cease to be accurate
(the limit of rank $N\rightarrow \infty$ does not commute with the limit
$\delta \rightarrow 0$, for $\alpha < 3$), but we are able to write
all the cumulants of  the width distribution distribution directly in
this limit (\sect{phisw}) via a subtle asymptotic expansion in powers
of $\delta$.  We prove that in the small-window limit the
width distribution becomes independent of the boundary condition. We also
show how to practically compute the ($\alpha$-dependent) universal width
distribution directly in this limit.  For $\alpha > 3$ (\sect{phiswag3}),
the problem simplifies. The characteristic function in the small-window
limit then corresponds to a matrix with only one non-zero eigenvalue,
and the corresponding width distribution no longer depends
on $\alpha$. We give its explicit form.  Finally, we compute the
logarithmic corrections to these asymptotic results for odd-integer
values of $\alpha$.  In \app{formula} and \app{generic_cumulant},  we
provide technical details of our calculation.

In a previous paper on the same subject \cite{rosso_raoul}, we already
studied the second moment of the width distribution, and presented
arguments for the universality in the small-window limit. The more
complete, and more concrete, calculations of the present paper rely
on the representation of the characteristic function in terms of the
spectrum of a matrix, which was not contained in \cite{rosso_raoul}.


\section{Average width} 
\slabel{avwidth}

In a self-affine (critical) system, the length $L$ of the total interval is, as mentioned,
the only characteristic length and the average value of the total
width scales as $\mean{w_2} \propto L^\zeta$ ($\mean{\ldots}$
denotes the ensemble average).

\subsection{Average width for $\delta=1$} 
\slabel{avwidth1}
Fixing $\delta=1$ and $x=0$ in \eq{w2_window} and integrating over $t$
yields
\begin{equation*}
   w_2= 
   \begin{cases}
   \frac{1}{2} \sum_{n=1}^\infty \glb a_n^2+b_n^2\grb & \text{(full
   periodic series)} \\ \sum_{n=1}^\infty
   \glc \frac{1}{2}s_n^2-\sum_{m=1}^\infty \frac{(1-(-1)^m)(1-(-1)^n)
   s_n s_m}{\mpi^2 n m} \right] & \text{(sine series)}\\ \frac{1}{2}
   \sum_{n=1}^\infty c_n^2 &\text{(cosine series)}.
   \end{cases}
\end{equation*}
From  the Gaussian probability distributions given in
\eq{action}-\eq{sigmasc}, the ensemble averages are 
\begin{equation*}
   \frac{\mean{w_2}}{L^{\alpha-1}}=
   \begin{cases}
   \frac{2}{(2 \mpi)^{\alpha}} \zeta(\alpha) & \text{(full periodic
   series)}\\ \frac{1}{ \mpi^{\alpha}} \zeta(\alpha)-\frac{2}{
   \mpi^{2}}\frac{2^{\alpha+2}-1}{(2 \mpi)^{\alpha}} \zeta(\alpha+2)
   &\text{(sine series)} \\ \frac{1}{ \mpi^{\alpha}} \zeta(\alpha) &
   \text{(cosine series)},
   \end{cases}
\end{equation*}
where $\zeta(x) = \sum_{n=1}^{\infty}1/n^x$ is the Riemann Zeta function.
For all $\alpha$, the average width of the full periodic series is smaller
than that of the cosine series, as is quite natural.  For $\alpha=2$,
the average widths of the full periodic series and the sine series
coincide, because the two differ only through boundary conditions for
the derivative of $u$, that a Markovian process is insensitive to.
\begin{figure}
   \centerline{
   \epsfxsize=7.0cm
   \epsfclipon
   \epsfbox{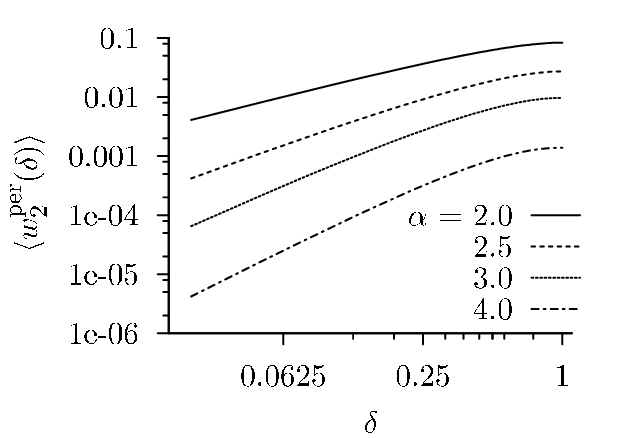}
   \epsfxsize=7.0cm
   \epsfclipon
   \epsfbox{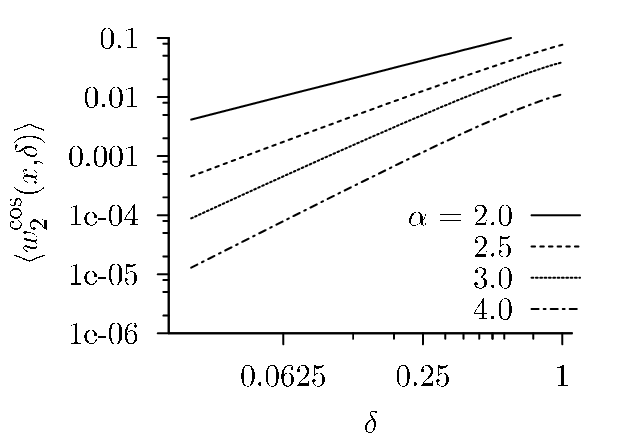}}
   \caption{Mean square width as a function of window width, for different values of $\alpha$,  
   for the full periodic series (\emph{left}) and the cosine 
   series (\emph{right}, with $x=\half(1 - \delta)$) (from
   \eqtwo{sum_w2_periodic}{sum_w2_free}).}  
   \flabel{mean_square_width}
\end{figure}

\subsection{Average width for $\delta<1$}
\slabel{avwidth2}
We now compute the average width for $\delta < 1$ for the full
periodic series, the cosine, and the sine series. The width
$w^{\period}_2(\delta)$ is independent of the origin $x$ but the
$x$-dependence of the average width cannot be neglected in the other
two cases.  For the full periodic series, one obtains from
\eqtwo{full_period_series}{w2_window}
\begin{equation}
   w^{\period}_2(\delta)= \sum_{n,m=1}^{\infty} a_n a_m C_{nm}(\delta)
   + a_n b_m I_{nm}(\delta) +b_n b_m S_{nm}(\delta),  
   \elabel{w2_expr}
\end{equation}
where the coefficients $C_{nm}(\delta)$ are given by elementary integrals:
\begin{equation}
   C_{mn}= \frac{1}{\delta} \int_x^{x + \delta} \dd{t} \cos( 2\mpi m t) \cos( 2\mpi
   n t) -\frac{1}{ \delta^2} \int_x^{x+\delta} \dd{t} \cos( 2\mpi m t) 
   \int^{ x + \delta}_x \dd{t} \cos(2\mpi n t).
   \elabel{cnm_coeff}
\end{equation}
Analogously, $S_{nm}$ and $I_{nm}$ can be expressed in terms of
sine--sine and cosine--sine integrals. For the full periodic series,
these coefficients are naturally independent of $x$.

\eqq{w2_expr} allows to compute $w^{\period}_2$ for one given
sample. Integrating over the Gaussian Fourier components
$\SET{a_n,b_n}$, we get
\begin{equation}
   \mean{w_2^{\period}(\delta)}= \frac{L^{\alpha-1}}{2^{\alpha-1}
   \mpi^{\alpha}} \sum_{n=1}^{\infty} \frac{C_{nn} +
   S_{nn}}{n^{\alpha}},\ \text{where} \quad C_{nn} + S_{nn}= 1 -
   \frac{1- \cos(2 \mpi n \delta )}{2 (\mpi n \delta)^2}.
   \elabel{sum_w2_periodic}
\end{equation}
The sum in \eq{sum_w2_periodic} is easily evaluated for finite $\delta$
(see \fig{mean_square_width}). The limit of this mean value for $\delta
\to 0$ cannot be obtained by a naive Taylor expansions of each term in
this infinite sum in \eq{sum_w2_periodic}, because it is not uniformally
convergent in the interval $\delta \in [0,1]$ for $\alpha<3$ (in the
limit $\delta \to 0$, the terms of the sum behave as $ (C_{nn}(\delta)
+ S_{nn}(\delta))n^{-\alpha}\sim \delta^2 n^{2-\alpha}$, producing a
diverging series for $\alpha<3$). As we will discuss in detail later, the
higher cumulants of the width distribution are given by multiple infinite
sums which present the same pathology as the sum in \eq{sum_w2_periodic}.
To sum the series in the limit $\delta \rightarrow 0$, we have derived
a very useful expansion formula which has the same structure as the
Euler-Maclaurin formula:
\begin{equation}
   \sum_{n=1}^{\infty} \frac{f(n\delta)}{n^{\alpha}} =
   \delta^{\alpha-1} \int_{0}^{\infty} \dd{t}
   \glb\frac{f(t)}{t^\alpha} - \sum_{m=0}^{\lfloor \alpha \rfloor-1}
   \frac{f^{(m)}(0) t^{m-\alpha}}{m!} \grb+ \sum_{m=0}^{\infty}
   \delta^m f^{(m)}(0) \frac{\zeta(\alpha-m)}{m!}, 
   \elabel{sum_integral_zeta}
\end{equation}
where $\lfloor \alpha\rfloor$ is the integer part of $\alpha$.
\eqq{sum_integral_zeta} holds inside the convergence radius $\delta=1$
for all the quantities considered in this work (the formula is 
proved in \app{formula}). For analytic functions
$f(z)$ and non-integer $\alpha$, the first term on the right can be
interpreted as the naive limit of the sum as $\delta \rightarrow 0$,
with $t=n\delta$, whereas the second term contains the Taylor
expansion of $f(n\delta)$ around zero. For integer $\alpha$, the
singularity of $\zeta(1)$ generates additional logarithms (see
\app{formula}, again). Using the expansion formula of
\eq{sum_integral_zeta}, one arrives at:
\begin{equation}
   \frac{\mean{w_2^{\period}(\delta)}}{L^{\alpha-1}} =
   \frac{2^{-\alpha-1}}{\zeta(-\alpha-1)}
   \frac{\zeta(\alpha+2)}{\mpi^{\alpha+2}} \delta^{\alpha-1} +
   \frac{4}{(2\mpi)^\alpha} \sum_{n=1}^\infty
   (-1)^{n+1}\frac{\zeta(\alpha-2n)}{(2 n +2)!}  (2 \mpi \delta)^{2n}.
   \elabel{w2_series}
\end{equation}
To take into account the logarithmic corrections for odd integer $\alpha$ we
must use \eq{formula_integer} instead of
\eq{sum_integral_zeta}. The final result writes:
\begin{equation*}
   \frac{\mean{w_2^{\period}(\delta)}}{L^{\alpha-1}}=(-1)^{\frac{\alpha+1}{2}}
   \frac{\psi_0(\alpha+2)-\log (2 \mpi \delta)}{2 \mpi
   (\alpha+1)!}\delta^{\alpha-1}+\frac{4}{(2\mpi)^\alpha}
   \sum_{n\ne\frac{\alpha-1}{2} } (-1)^{n+1}\frac{\zeta(\alpha-2n)}{(2
   n +2)!}  (2 \mpi \delta)^{2n}\quad (\text{$\alpha$ odd integer}),
\end{equation*}
where $\psi_0(z)$ is the digamma function. An expansion analogous
to \eq{w2_series}
appears in the correlation function governing the density of
zero-crossings of a Gaussian function \cite{majumdar_bray}. For
integer even $\alpha$, the series in \eq{w2_series} is finite (because
$\zeta(x)=0$ for $x=\SET{-2,-4,\dots}$).  The mean square widths for
the periodic random walk ($\alpha=2$) and the driven curvature model
($\alpha=4$) then have compact expressions:
\begin{equation*}
   \frac{\mean{w_2^{\period}(\delta)} }{L^{\alpha-1}}= 
   \begin{cases}
   \frac{1}{6}\delta - \frac{1}{12}\delta^2 &\text{for $\alpha=2$}\\
   \frac{1}{144}\delta^2- \frac{1}{120}\delta^3 +\frac{1}{360}\delta^4
   & \text{for $\alpha=4$}
   \end{cases}.
\end{equation*}

For the cosine series, we obtain, from \eqtwo{cosine_series}{w2_window}, 
the mean squared width
\begin{equation}
   w_2^{\cos}(x,\delta)= \sum_{n,m=1}^{\infty} c_n c_m
   D_{nm}(x,\delta), 
   \elabel{w2_free_expr}
\end{equation}
where the coefficients are given by a symmetric matrix
\begin{equation}
   D_{mn}(x, \delta)= \frac{1}{\delta} \int^{x+\delta}_x \dd{t}  \cos( \mpi m
   t) \cos( \mpi n t)  -\frac{1}{ \delta^2} \int^{x+\delta}_x \dd{t} 
   \cos( \mpi m t) \int^{x+ \delta}_x \dd{t} \cos(\mpi n t).
   \elabel{dnm_coeff}
\end{equation}
$D_{nm}$ is the overlap matrix of the basis functions in \eq{cosine_series} on the
interval $[x,x+\delta]$. For $\delta=1$, the basis functions are, by 
construction, an
orthonormal set, and $D_{nm}=\half \delta_{nm}$ .

Integrating over the Gaussian variable $\SET{c_n}$, the average width
$\mean{w_2^{\cos}}$ becomes
 \begin{equation}
   \mean{w_2^{\cos}(x,\delta)}= \frac{ 2 L^{\alpha-1}}{\mpi^{\alpha}}
   \sum_{n=1}^{\infty} \frac{D_{nn}(x,\delta)}{n^{\alpha}}.
   \elabel{sum_w2_free}
\end{equation}
The sum in \eq{sum_w2_free} is again easily evaluated, for any value of $x$
and $\delta$. The behavior of $\mean{w_2^{\cos}(x,\delta)}$ as a function of
$\delta$ is shown in \fig{mean_square_width} for $x=\half (1-\delta)$. For small
$\delta$ we notice that $\mean{w_2^{\cos}(x,\delta)}\propto \delta^{\alpha -
  1}$ for $\alpha < 3$ and $\mean{w_2^{\cos}(x,\delta)}\propto \delta^2$ for
$\alpha > 3$.  A special case is
again $\alpha =2$ where we have $\mean{w_2^{\cos}(x,\delta)} \propto \delta$
for all $\delta$.

\subsection{Small-window limit}
\slabel{avsw}

In the following, we extract the universal behavior of
$\mean{w_2^{\cos} ( x, \delta)}$ for a small window $[x,x+\delta]$
far from the boundaries (for $\delta \ll \min(x,1-x)$).  For
$\alpha<3$, \eq{sum_w2_free} can be expanded as
\begin{equation}
   \mean{w_2^{\cos}(x, \delta)}=\frac{ 2 L^{\alpha-1}}{\mpi}
   \delta^{\alpha-1}\int_0^{\infty} \dd{t}
   \frac{D(t,x/\delta)}{t^{\alpha}}+O(\delta^{2}),
   \elabel{w2_free_dominant_m3}
\end{equation}
where the function $D(t,x/\delta)$ is defined through $D(\mpi n
\delta,x)=D_{nn}(x)$.  We write $D(t,x/\delta)$ as a sum of two terms:
\begin{equation*}
   D(t,x/\delta)=h(t,x/\delta)+\half\glc C(t)+S(t)\grc. 
\end{equation*}
The function
$h(t,x/\delta)$ contains all the $x$-dependence of $D(t,x/\delta)$:
\begin{equation*}
   h(t,\frac{x}{\delta})=\cos(2t(\frac{x}{\delta}+\frac{1}{2}))
   \glc \frac{\sin(2t)}{4t}-\frac{1}{t^2}\sin(\frac{t}{2})^2 \grc.
\end{equation*}
For $x/\delta \gg 1$, $h(t,x/\delta)$ oscillates rapidly and gives a
vanishing contribution to the integral in \eq{w2_free_dominant_m3}. In
particular, we can show that
\begin{equation*}
   \lim_{x/ \delta \to \infty} \frac{\mean{w_2^{\cos}(x,
   \delta)}}{L^{\alpha-1}} =\frac{2^{-\alpha-1}}{\zeta(-\alpha-1)}
   \frac{\zeta(\alpha+2)}{\mpi^{\alpha+2}} \delta^{\alpha-1}+O \glc
   \expb{-x/\delta} \grc.
\end{equation*}
Comparing the above expression with \eq{w2_series}, we conclude that, for
$\alpha <3$, the dominant contributions to $\mean{w_2^{\cos}(x,\delta)}$
and to $\mean{w_2^{\period}(\delta)}$ coincide for small $\delta$
(compare also with \fig{mean_square_width}).  This conclusion can be
extended to the sine series, where the explicit $x$-dependence is again
due to oscillatory terms that vanish in the small-window limit.

As mentioned before, the series in \eq{sum_w2_free} is uniformly convergent for
$\alpha >3$, so that one can take the limit $\delta \to 0$ for each mode and then
do the sum. Using the expansion 
\begin{equation}
 D_{nm}(x)=\frac{1}{12}nm \mpi^{2}\sin(\mpi m x)\sin(\mpi n x)
   \delta^2+O(\delta^4) 
   \elabel{dnm_expan_coeff}
\end{equation}
in \eq{sum_w2_free}, we obtain
\begin{equation}
   \frac{\mean{w_2^{\cos}(x, \delta)}}{L^{\alpha-1}} = \frac{1}{6}
   \lb \sum_{n=1}^{\infty} \frac{\sin^2( n \mpi x)}{n^{\alpha -2}} \rb
   \mpi^{2-\alpha} \delta^2 + O(\delta^{\alpha-1}, \delta^4)
   \elabel{w2_free_series_g3}.
\end{equation}
For $x=0$ and $x=1$ the $\delta^2$ term vanishes,  while for
$0<x<1$ it is smaller than the corresponding term for the periodic
case. This behavior is consistent with the fact that, for $\alpha>2$,
the cosine series imposes vanishing derivatives at the end points. Hence
these boundary conditions force $\mean{w_2^{\cos}} $ to be smaller than
$\mean{w_2^{\period}}$ for all $0<x<1$. The sine series gives a result
analogous to the one for the cosine series. It suffices to replace the
sine in the sum of \eq{w2_free_series_g3} by a cosine.

\section{Width distribution}
\slabel{phi}
 
The scaling behavior of the average width gives access to the value of
$\alpha$ and thus to the roughness exponent $\zeta=(\alpha-1)/2$. However,
the value of this exponent only depends on the $2$-point correlation
functions, and captures no finer geometric properties of the  function
$u(t)$. To discriminate between a Gaussian and a non-Gaussian
function, one must have access to higher cumulants, as they are
contained in  the sample-to-sample fluctuations of the two-point
correlation functions. The distribution $\PCAL(w_2)$ has been 
used to analyze numerical and experimental data \cite{rosso_width2,
bramwell.xy, marinari_racz, vandembroucqphi, zapperi}.
The width distribution also allows to estimate the roughness
exponent from experimental data which are not sufficiently good to plot 
$\mean{w_2(\delta)}$ vs $\delta$ over several orders of magnitude in $\delta$.

For general values of
$\alpha$, the width distribution has been computed for the
entire interval ($\delta=1, x=0$), for the full periodic series,
\cite{racz.random.94,plischke.curvature.94,antal.1overf.01}, and for
the cosine series \cite{rosso_width2}.  The underlying simplification
with respect to the calculations in the present paper is 
that the matrix
of coefficients for $\delta=1$ satisfies $I_{nm}=0$ while $C_{nm}$, $S_{nm}$ and $D_{nm}$
are diagonal, as evident in \eqtwo{cnm_coeff}{dnm_coeff}.  In order to
study window boundary conditions (but also for the sine series),
the previous framework must be generalized to non-diagonal matrices.

The width distribution can be obtained from the symmetric matrices
\begin{equation}
   A^{\cos}_{nm}= 2 \sigma^{\cos}_n D_{nm} \sigma^{\cos}_m.
   \elabel{A_fr}
\end{equation}
and 
\begin{equation}
   A^{\period} =\glb \begin{array}{cc}
   2\sigma_n^\period C_{nm}\sigma^\period_{m} & \sigma_n^\period I_{nm}\sigma^\period_{m} \\ 
   \sigma_n^\period I_{mn}\sigma_m^\period & 2
   \sigma^\period_n S_{nm} \sigma^\period_{m} \end{array} \grb, 
   \elabel{A_per}
\end{equation}
respectively.
Concretely (see \eqtwo{w2_expr}{w2_free_expr}), individual realizations
of the mean square width distribution are generated by multiplying the
matrices $A$ in \eqtwo{A_fr}{A_per} by vectors of normal distributed
Gaussian variables.  This allows to obtain $\PCAL(z= w_2/\mean{w_2})$ approximately
through direct simulation (see \cite{werner_book}).

On the other hand, one can compute all the moments of the distribution
from contractions of a given matrix $A$ (which can stand for $A^{\cos}$,
$A^{\sin}$, or $A^{\period}$).  For example, the second moment of the
width distribution is given by
\begin{equation}
   \mu_2 = \mean{w_2^2} = \half \sum_{n,  m} A^{\cos}_{nm} A^{\cos}_{mn} + \frac{1}{4} \glb \sum_{n} A^{\cos}_{nn} \grb^2
   = \half  \Tr \glc A^{\cos} \grc^2 + \glc  \half \Tr A^{\cos} \grc^2.
   \elabel{trace_naive}
\end{equation}
Similar expressions exist for higher moments of the width distribution.  As
shown in \app{generic_cumulant}, the cumulant $\kappa_l$ of the rescaled width
distribution can be expressed in a simpler way than the moments, as a trace of
the matrix $A$ taken to the $l$th power:
\begin{equation*}
   \kappa_{l}=\frac{(l-1)!}{2\mean{w_2}^{l}}\Tr 
   \glc A^{l} \grc
   \elabel{cumulant_expr_trace}
\end{equation*}
This is already apparent in \eq{trace_naive}, where $\mu_2 =  \mean{w_2}^2 (\kappa_2 + \kappa_1^2$).
The cumulants are thus given in terms of the eigenvalues 
$\SET{\lambda_1,\lambda_2, \dots}$  
of the matrix $A$ as $\Tr A^l = \sum_k \lambda_k^l$:
\begin{equation}
   \kappa_{l}=\frac{(l-1)!}{2}\frac{\sum_k \lambda_k^l}{(\sum_k \lambda_k)^l}.
   \elabel{cumulant_expr_eigen}
\end{equation}
The cumulants in \eq{cumulant_expr_eigen}
yield an explicit formula for the cumulant-generating
function $\Psi(s/\mean{w_2})$, where
\begin{equation*}
   \Psi(s) = \sum_{k=1}^\infty \sum_{l=1}^\infty \frac{\lambda_k^l}{2 l } s^l = 
   \sum_k \half \log\glb 1 - \lambda_k s\grb.
\end{equation*}
The characteristic function $f(s)= \exp(\Psi(\mi s))$, the exponential of the cumulant-generating
function $\Psi(\mi s)$, is given by
\begin{equation}
   f(s)= \prod_{k} \frac{1}{\sqrt{1 -\mi \lambda_k s}}.
   \elabel{characteristic_expr}
\end{equation}
The spectrum of the matrices under consideration is such that the infinite
product in \eq{characteristic_expr} is uniformally convergent.  The associated width
distribution is recovered through an inverse Fourier transform, 
\begin{equation}
   \PCAL(z= w_2/\mean{w_2})  = \frac{1}{2 \mpi} \int_{-\infty}^{\infty} \dd{s} \exp(-\mi z s)
   f(s), 
   \elabel{inverse_fourier}
\end{equation}
which can be obtained by straightforward Riemann integration
because the branch points $s_k = \mi /\lambda_k$ are away from the real axis.  
\subsection{Width distribution for  the entire interval ($\delta=1$)}
\slabel{phiint}
As mentioned above, in the case $\delta=1$, the matrices 
\begin{equation}
   A^{\period}=\glb \begin{array}{cc}
   \glc \sigma_n^\period \grc ^2 \delta_{nm}& 0\\ 
   0 & 
   \glc \sigma^\period_n \grc ^2 \delta_{nm}\end{array} \grb,  \quad \mbox{and}\quad 
   A^{\cos}= \glc \sigma_n^{\cos} \grc ^2\delta_{nm},
   \elabel{amatrix_d1}
\end{equation}
are diagonal and the computation of the cumulants from \eq{cumulant_expr_eigen} is direct: 
\begin{equation*}
   \kappa_l=
   \begin{cases}
   (l-1)!\zeta(l\alpha)/\zeta(\alpha)^l &\text{(full periodic
   series)}\\ (2l-2)!!\zeta(l\alpha)/\zeta(\alpha)^l &\text{(cosine
   series)}.
   \end{cases}
\end{equation*}
The cumulants for the sine series are more complicated because the
corresponding matrix $A^{\sin}$ presents non-diagonal terms coming from the
non-vanishing mean $\frac{1}{L}\int_{0}^{L} \dd{t} \sinb{\mpi nt/L}$ for odd values of
$n$. The second cumulant, for example, is given by
\begin{equation*}
   \kappa^{\text{sin}}_2=\frac{2\zeta(2\alpha)+ 2^{-2\alpha+1}\glb
   16\mpi^{-4}(2^{\alpha+2}-1)^2\zeta(\alpha+2)^2
   -8\mpi^{-2}(2^{2\alpha+2}-1) \zeta(2\alpha+2)\grb}{\glc
   \zeta(\alpha)- 2^{-\alpha+1}\mpi^{-2}(2^{\alpha+2}-1)
   \zeta(\alpha+2)\grc ^2}.
\end{equation*}
$\kappa^{\text{sin}}_2$ agrees with $k^{\period}_2$ only for
$\alpha=2$, as expected.
In the case of the full periodic and the cosine series,  the characteristic functions
assume a simple form 
\begin{equation*}
   f(s)=
   \begin{cases}
   \prod_{k} (1 -\mi L^{\alpha-1}(2\mpi )^{\alpha} k^{-\alpha} s)^{-1} &\text{(full periodic series)}\\
   \prod_{k} (1 -\mi  L^{\alpha-1} \mpi^{\alpha} k^{-\alpha} s)^{-\half} &\text{(cosine series)}.
   \end{cases}
\end{equation*}
As already discussed in \cite{antal.1overf.01}, the two-fold
degeneracy of the spectrum  of $A^{\period}$ for $\delta=1$, evident in
\eq{amatrix_d1}, yields a characteristic function $f(s)$ with simple poles
on the imaginary axis. This simplifies the inverse Fourier transformation.

\subsection{Width distribution for windows (finite $\delta$)}
\slabel{phi_finite_delta} 

The expressions allowing to recover the width distribution from the
eigenvalues of the matrices $A^{\cos}, A^{\period}$, and $A^{\sin}$
remains valid for intervals $\delta<1$, even though the computer  must
now be used for finite approximations of these infinite matrices, which
are no longer diagonal.
These calculations can be easily checked
by direct simulation, as mentioned before. The outcome of this
analysis is shown in \fig{width_fluctuation} for the case $\alpha=5/2$.
\begin{figure}
   \centerline{
   \epsfxsize=7.0cm
   \epsfclipon
   \epsfbox{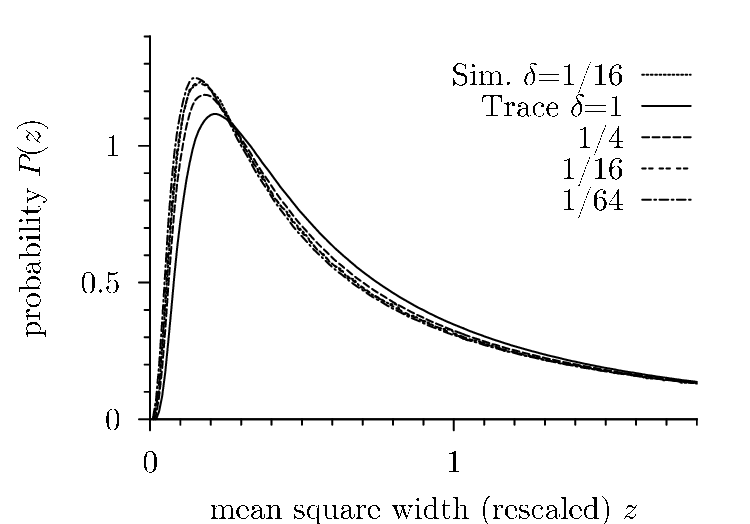}
   \epsfxsize=7.0cm
   \epsfbox{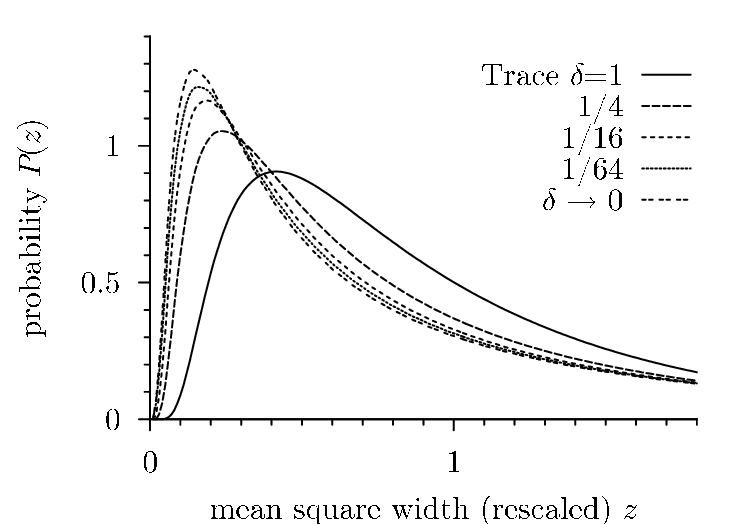}}
   \caption{Rescaled probability distributions for the mean square width for $\zeta=0.75$ ($\alpha=2.5$)
   for the cosine series (\emph{left}) and the full periodic series (\emph{right}), obtained from 
   \eqtwo{characteristic_expr}{inverse_fourier}. 
   The result of direct simulations for $\delta=1/16$ (from \eq{w2_free_expr}), and the 
   solution in the  limit $\delta \rightarrow 0$ (from the matrix in \eq{double_tilde_A}) are also shown. 
   Sizes of matrices are
   $N=512$ and $N=1024$.}
   \flabel{width_fluctuation}
\end{figure}
It is evident that the width distribution changes with the sample
size for the value of $\alpha $ chosen. Furthermore, the direct evaluation
of the rescaled width distribution for finite but small $\delta$
suggests that this distribution becomes universal (independent of the
boundary conditions) in the limit $\delta \rightarrow 0$. This scenario is
confirmed by comparing the evolution of the (normalized) spectra of $A^{\period}$ and
$A^{\cos}$, as shown in \fig{spectra_075_2.5}.

We then compute $\PCAL(z= w_2/\mean{w_2})$ directly in the limit
$\delta\to 0$.  However, for small $\delta$,  increasingly larger matrices
$A^{\period}$ and $A^{\cos}$ must be considered because of the non-uniform
convergence  of the traces of these matrices for $\delta \in [0,1]$.

\begin{figure}
   \centerline{
   \epsfxsize=7.0cm
   \epsfclipon
   \epsfbox{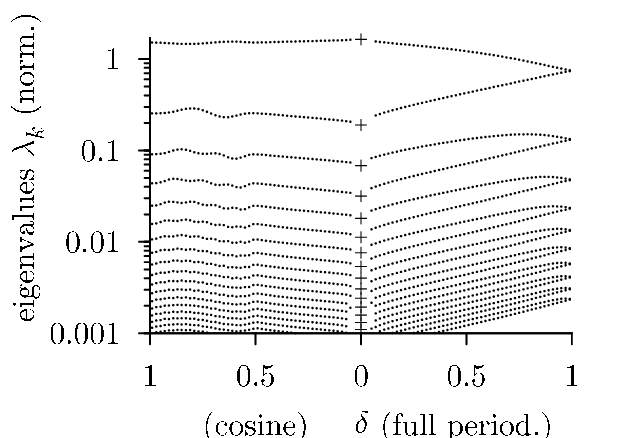}}
   \caption{Eigenvalues $\SET{\lambda_1,\lambda_2,\dots}$ for the
   cosine and the full periodic series as a function of window size
   $\delta$ ( $\alpha=2.5$ ($\zeta=0.75$), for the cosine series,
   $x=\half(1-\delta)$). The spectrum in the small-window limit $\delta
   \rightarrow 0$ (\emph{crosses}), is obtained in \sect{phisw}. Eigenvalues
   are normalized as $\sum_k \lambda_k=2$.  }
   \flabel{spectra_075_2.5}
\end{figure}

In \sect{phisw} we determine directly the width distribution at the
limit $\delta \to 0$. In particular, analogously to the computation
of asymptotics of the average width (\sect{avsw}), the expansion
formula \eq{sum_integral_zeta} is proven useful to compute the
cumulants of the distribution in the small-window limit for $\alpha<3$
(\sect{phiswam3}). This analysis serves two purposes: it proves the
universality of the width distribution and provides a high-precision
method to compute it directly in the limit $\delta \rightarrow 0$.

\subsection{Width distribution in the small-window limit}
\slabel{phisw} 

In \sect{avsw} we computed the mean square widths
 $\mean{w_2^{\period}(\delta)}$ and $\mean{w_2^{\cos}(x,\delta)}$ in
 the small-window limit.  We now determine the rescaled width distribution
 $\PCAL(z)$ in this limit.

\subsubsection{Small-window limit ($\alpha < 3$)}
\slabel{phiswam3}

The computation of $\kappa_{l}$ as a $\delta$-expansion is along the lines of
the previous determination of $\mean{w_2}$. For $\alpha < 3$, we handle 
the non-uniform convergence of the traces in \eq{cumulant_expr_eigen} for
$\delta\to 0$ using the expansion formula \eq{sum_integral_zeta}.
The leading contribution to $\kappa_{l}$, $l=1,2,\dots$,  is given by
multiple integrals:
\begin{gather}
   \kappa^{\period}_l=\frac{\mpi^{l(\alpha - 1) }(l-1)!}{2 \widehat{w}_2^l}
   \int_{0}^{\infty} \cdots \int_{0}^{\infty}
   \prod_{n=1}^{l}\frac{\ddd{t_{n}}}{t_n^\alpha} \Tr \glc A^{\period}(t_1,t_2)
   A^{\period}(t_2,t_3)\cdots
   A^{\period}(t_l,t_1) \grc +O(\delta^{3-\alpha}) \elabel{kappa_asymp_per}\\
   \kappa^{\cos}_l(x/\delta)=\frac{\mpi^{l(\alpha - 1) }(l-1)!}{2 
      \widehat{w}_2 ^l} \int_{0}^{\infty} \cdots \int_{0}^{\infty}
   \prod_{n=1}^{l}\frac{\ddd{t_{n}}}{t_n^\alpha} \glc A^{\cos}(t_1,t_2,\frac{x}{\delta})
   A^{\cos}(t_2,t_3,\frac{x}{\delta})\cdots
   A^{\cos}(t_l,t_1,\frac{x}{\delta}) \grc +O(\delta^{3-\alpha}).
   \elabel{kappa_asymp_cos}
\end{gather}
In the above equations, $\widehat{w}_2$ is the prefactor of the 
leading term in the expansion of $\mean{w_2(\delta, x)}$, given 
in \eq{w2_series}:
\begin{equation*}
\widehat{w}_2 =\frac{2^{-\alpha-1}}{\zeta(-\alpha-1)}
   \frac{\zeta(\alpha+2)}{\mpi^{\alpha+2}}, 
\end{equation*}
which is independent of the boundary
conditions. $A^{\cos}(t_n,t_{n+1},x/\delta)$ and
$A^{\period}(t_n,t_{n+1})$ are obtained by replacing $\mpi n \delta$ with
$t_n$ in $A^{\cos}$ and in $A^{\period}$. We now verify for each
cumulant
\begin{equation}
   \kappa_{l}^{\cos}(x/\delta \rightarrow \infty) \rightarrow
   \kappa_{l}^{\period} +O( e^{-x/\delta}).  
   \elabel{equivalence}
\end{equation}
Extracting the $x/\delta$ independent part of $\kappa^{\cos}_l$ one obtains 
\begin{equation}
   \kappa^{\cos}_l=\frac{\mpi ^{l(\alpha - 1) }(l-1)!}{2 \widehat{w}_2^l}
   \int_{0}^{\infty} \cdots \int_{0}^{\infty}
   \prod_{n=1}^{l}\frac{\ddd{t_{n}}}{t_n^\alpha} \Tr \glc \tilde{A}^{\cos}(t_1,t_2)
   \tilde{A}^{\cos}(t_2,t_3)\cdots
   \tilde{A}^{\cos}(t_l,t_1) \grc + \text{oscill. terms}, 
   \elabel{kappa_asym_cos_smooth}
\end{equation}
where $\tilde{A}^{\cos}$ is now independent of $x$ and given by 
\begin{equation}
   \tilde{A}^{\cos}(t,t') =\frac{4 L^{\alpha-1}}{\mpi^\alpha} 
   \glb \begin{array}{cc} \frac{1}{2(t-t')}
   \sina{\frac{t-t'}{2}}-\frac{1}{tt'}
   \sina{\frac{t}{2}}\sina{\frac{t'}{2}} & \frac{1}{2(t+t')}
   \sina{\frac{t+t'}{2}}-\frac{1}{tt'}
   \sina{\frac{t}{2}}\sina{\frac{t'}{2}} \\ \frac{1}{2(t+t')}
   \sina{\frac{t+t'}{2}}-\frac{1}{tt'} \sina{\frac{t}{2}}
   \sina{\frac{t'}{2}} & \frac{1}{2(t-t')}
   \sina{\frac{t-t'}{2}}-\frac{1}{tt'}
   \sina{\frac{t}{2}}\sina{\frac{t'}{2}}   
   \end{array} \grb .
   \elabel{A_limit_no_rescaling}
\end{equation}
Now it is straightforward to verify that the
matrices $A^{\period}$ and  $A^{\cos}$ satisfy
\begin{equation*}
   \Tr\glc \tilde{A}^{\cos}(t_1,t_2)\tilde{A}^{\cos}(t_2,t_3)\cdots
   \tilde{A}^{\cos}(t_l,t_1)  \grc =\Tr \glc   A^{\period}(t_1/2,t_2/2)
   A^{\period}(t_2/2,t_3/2)\cdots A^{\period}(t_l/2,t_1/2)  \grc .
\end{equation*}
This establishes the validity of \eq{equivalence}.

Closed analytic expression have not been obtained for the above integrals.
Actually the problem of computing the integrals in \eq{kappa_asym_cos_smooth}
reduces to the solution of a homogeneous Fredholm equation of the second kind:
\begin{equation*}
   \int_{0}^{\infty} \dd{t}  \sum_{j=1}^2 \tilde{A}^{\cos}(t,t')_{ij} g_{j}(t')=
   \lambda g_{i}(t), 
\end{equation*}
where the $2\times 2$ matrix  
$\tilde{A}^{\cos}(t,t')$ is a compact kernel with a discrete set
of eigenvalues $\lambda_k$ converging to $0$, which encode all the information 
on the cumulants $\kappa_l$, in the same way as for the discrete case (\eq{cumulant_expr_eigen}).

The spectrum of the kernel $\tilde{A}^{\cos}(t,t')$ 
is most easily obtained by discretizing the variables $t$ and $t'$ 
on an equally spaced grid with $t_k = \Delta (k-\half),\ k=1\TO N$,  with an upper
cut-off for the integrations. However, the singularities of the  integrands makes the convergence rather
slow
($\sim t_n^{2 - \alpha}$ for $t_n \rightarrow 0$ (with all other variables kept finite)).
The divergence in the integrals at small $t$ is eliminated by 
a standard change of variables: an integral of a function 
diverging as $1/t^{\gamma}$ for $t \rightarrow 0$ can be written as
\begin{equation*}
   \int_0^\infty \dd{t} f(t) = \frac{1}{1 - \gamma} \int_0^{\infty} \dd{t} t^{\gamma/(1-\gamma)}
   f\glc t^{1/(1-\gamma)} \grc.
\end{equation*}

\begin{figure}
   \centerline{
   \epsfxsize=7.0cm
   \epsfclipon
   \epsfbox{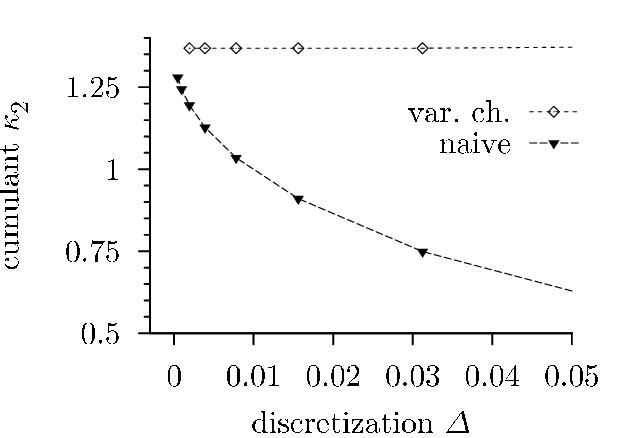}}
   \caption{Cumulant $\kappa_2$ of the rescaled width distribution for 
   $\alpha=2.5$ 
   from a naive discretization of the integral in \eq{kappa_asym_cos_smooth}, 
   and after the change of variables in \eq{double_tilde_A}
   (Riemann integration in the interval $t \in [0,15]$, using $\Delta=t_{i+1}-t_i$). }
   \flabel{convergence}
\end{figure}
The integrand on the right is constant for small $t$. 
Concretely, the change of variables in $\tilde{A}^{\cos}$
leads to a matrix 
\begin{equation}
   \tilde{\tilde{A}}^{\cos}(t,t')= \frac{1}{3-\alpha} t^{\frac{\alpha-2}{6 - 2 \alpha}}
   A^{\cos}(t^{1/(3-\alpha)}, t'^{1/(3-\alpha)}) t'^{\frac{\alpha-2}{6 - 2 \alpha}}
   \elabel{double_tilde_A},
\end{equation}
which can again be discretized.
The characteristic function of the width distribution is computed
from the spectrum of the $2N \times 2N$ 
matrix $\tilde{\tilde{A}}^{\cos}$ as discussed before.
To show what is gained by rescaling the matrix $\tilde{A}$ in 
\eq{A_limit_no_rescaling}, 
we have computed the second cumulant $\kappa_2$ on an equally spaced grid
with both versions (see \fig{convergence}). 
The rescaled matrices converge exceptionally well
with the discretization parameter $\Delta$.  In \fig{spectra_075_2.5}, the spectrum
of the rescaled kernel (for $\delta \rightarrow 0$) 
is compared to the spectra of $A^{\period}$
and $A^{\cos}$ for finite $\delta$. Only in the special case
$\alpha=2$ is the spectrum of the cosine series independent of $\delta$, because
of the Markov-chain property of the random walk.
For all $\alpha \ne 2$, the spectrum, thus the cumulants of the width distribution,
depend of $\delta$, but satisfy
\begin{equation*}
   \kappa_l^{\period}(\delta \rightarrow 0) = \kappa_l^{\cos}(\delta
   \rightarrow 0) \ne \kappa_l^{\cos}(\delta=1),
\end{equation*}
in a way illustrated  in \fig{spectra_075_2.5}.  As was already shown in
\fig{width_fluctuation} for $\alpha=2.5$, the convergence of the width
distribution for finite $\delta$ towards the asymptotic width distribution
is quite fast, even though the rate of convergence depends on the value
of $\alpha$.

\subsubsection{Small-window limit ($\alpha \ge 3$)}
\slabel{phiswag3}

For $\alpha \ge 3$, the integral term in \eq{sum_integral_zeta} is
sub-dominant, and the naive expansion of the matrices $A^\period$ and $A^{\cos}$ in powers
of $\delta$ becomes correct in the limit $\delta \rightarrow 0$.  
Using the expansions in \eq{dnm_expan_coeff} and, furthermore,  
\begin{gather}
   C_{mn}=O(\delta^4),\;\;S_{mn}=\frac{1}{3} m n \mpi^2
   \delta^2+O(\delta^4),\;\; I_{nm}=O(\delta^3),
\end{gather}
we can construct the matrices $A^{\period}$ and
$A^{\cos}$, and check that they have only a single non-zero eigenvalue. 
For illustration, we show in
\fig{spectrum_cos_periodic_3.5} the spectrum  of these two matrices  
for  $\alpha=7/2$ at finite $\delta$
(eigenvalues are normalized so that $\sum_k \lambda_k=2$).

\begin{figure}
   \centerline{
   \epsfxsize=7.0cm
   \epsfclipon
   \epsfbox{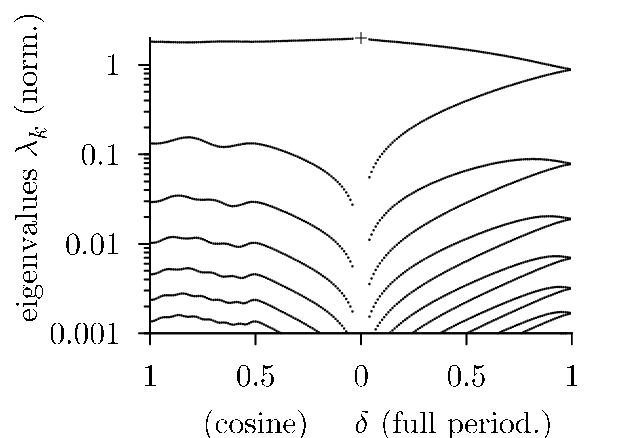}}
   \caption{Eigenvalue spectrum of $A^{\cos}$ and $A^{\period}$ for
   $\alpha=3.5$ ($\zeta=1.25$).  A single eigenvalue remains non-zero
   in the limit $\delta\rightarrow 0$ (\emph{cross}). 
   The normalization condition $\lambda_1 = \sum_k \lambda_k=2$ corresponds to
   the rescaling of the width distribution, with $\mean{w_2}=1$.
   }
   \flabel{spectrum_cos_periodic_3.5}
\end{figure}

From \eq{cumulant_expr_eigen} we then obtain  
\begin{equation*}
   \kappa_l^{\cos}(\delta \rightarrow 0)=\kappa_l^{\period}(\delta
   \rightarrow 0) =
   \begin{cases} (2l-2)!! + O(\delta^{\alpha
   -3}) & \text{for}\ \alpha > 3\\ (2l-2)!! + O(1/\log \delta)
       & \text{for}\ \alpha = 3
   \end{cases}.
\end{equation*}
The associated characteristic function reads:
\begin{equation*}
   f(s)=(1-2\mi \lambda s)^{-\half},
\end{equation*}
and its inverse Fourier transform gives the universal distribution  $\PCAL(z)$
for $\alpha > 3$,:
\begin{equation*}
   \PCAL(z)  = \frac{\expl{- z/2}}{\sqrt{2 \mpi z}}. 
\end{equation*}
By comparing the universal distribution for $\alpha>3$ with the ones with
$\alpha<3$ (see \fig{width_fluctuation}), one notices that the additional 
eigenvalues (and branch points contributing to the
characteristic function) allow the distribution $\PCAL(z)$ to vanish at
$z=0$, thus producing a local maximum in this function.

\section{Conclusions}
In conclusion, we have studied the geometric properties of functions with
a particularly simple expression in Fourier space: independent Gaussian
random variables.  We have restricted ourselves to one-dimensional
self-affine functions (characterized by a single length scale),
but our analysis evidently carries over to functions with more than
one length scale, and to higher dimensions. In real space, the geometrical
properties are non-trivial, and the boundary conditions play an important
role. This comes about because the real-space action, for all non-even
$\alpha$, contains non-local operators and, for all $\alpha \ne 2$,
is non-Markovian.

We have provided a simple and compact framework for studying the boundary
effects for general non-Markovian Gaussian processes by  relating the
characteristic function of the width distribution to the spectrum of
boundary-dependent infinite matrices, which essentially encode the overlap
of the basis functions of the Fourier series. The choice of Fourier basis
(resulting in the full periodic, cosine and sine series) determines the
different boundary conditions.  We have carried out a complete analysis
of the spectrum of these matrices for different values of $\alpha$
for the case of the function on the entire interval and also for the
function restricted to a window. The associated width distributions
could all be determined by solving for the eigenvalues of a matrix,
and by performing a straightforward inverse Fourier transform.

We have shown that the non-Markovian action propagates the effects of
boundary conditions over the entire interval. However, in the small-window
limit, the width distribution becomes universal (independent of boundary
conditions).  For $\alpha<3$, we showed how to compute the cumulants
of the width distribution in this limit, avoiding problems related to
the non-uniform convergence of the Fourier series.  For $\alpha>3$,
the problem of finding the universal width distribution drastically
simplifies, and we were able to write it down explicitly. Finally, we
have obtained the logarithmic corrections in the case of odd integer
$\alpha$, in particular for $\alpha=3$.

We hope that our work will be useful for the analysis of experimental
data (which usually correspond to our window boundary conditions, often
in the regime $\delta \rightarrow 0$, which we found to be universal). In
many experiments, the roughness exponent cannot be extracted reliably by
extrapolation, and the width distribution may provide crucial additional
information.

Computer programs that compute the width distributions for any value of
$\alpha$, both at finite $\delta$ and in the limit $\delta \rightarrow
0$, are  available \cite{foot_computers}.

\acknowledgments
The authors thank the Kavli Institute for Theoretical Physics, Santa
Barbara, for its hospitality.  This research was supported in part by
the National Science Foundation under the Grant No. PHY99-07949.

\appendix

\section{Expansion formula}
\slabel{formula}
In this appendix we derive the expansion formula of
\eq{sum_integral_zeta}, first for non-integer, then for integer values
of $\alpha$.

\subsection{Expansion formula for non-integer $\alpha$}
We consider the sum
\begin{equation*}
   \sum_{n=1}^{ \infty} \frac{f(n\delta)}{n^{\alpha}},
\end{equation*} 
where $f(z)$ is assumed to be a general analytic function. In order to
obtain the expansion in powers of $\delta$ of the above sum, we expand
$f(z)$:
\begin{equation*}
   \sum_{n=1}^{\infty}
   \frac{f(n\delta)}{n^{\alpha}}=\sum_{n=1}^{\infty}\sum_{m=0}^{\infty} \frac{
   f^{(m)}(0) }{m!\,n^{\alpha-m}} \delta^m.
\end{equation*}
For $\alpha-m<1$,  the sum over $n$ converges, and can
be directly evaluated using the Riemann Zeta
function $\zeta(\alpha-m)=\sum_n 1/n^{\alpha-m}$:
\begin{equation}
   \sum_{n=1}^{\infty} \frac{f(n\delta)}{n^{\alpha}}=\sum_{m=0}^{[\alpha]-1}
   \frac{ f^{(m)}(0) \zeta(m-\alpha) }{m!} \delta^m + 
   \sum_{n=1}^{\infty} \frac{\tilde{f}(n\delta)}{n^{\alpha}},
   \elabel{cut_sum_appendix}
\end{equation}
where $[\alpha]$ is the integer part of $\alpha$, and
$\tilde{f}(z)=\sum_{[\alpha]}^\infty f^{(m)}(0) z^m /m!$.  The second
term on the right hand side of \eq{cut_sum_appendix} can be expressed
as a contour integral in the complex plane:
\begin{equation}
   \sum_{n=1}^{\infty}
   \frac{\tilde{f}(n\delta)}{n^{\alpha}}=\frac{\delta^{\alpha-1}}{2
   \mi}\int_{C}\dd{z} \frac{\tilde{f}(z)}{z^{\alpha}}\cotb{\frac{\mpi}{\delta}z}, 
   \elabel{int_su_C}
\end{equation}
where the contour $C$ encircles the poles $z_n=n\delta$, $n=1,2\dots$ of
the function $G(z)=[\tilde{f}(z)/z^{\alpha}]\cotb{\mpi z/ \delta} $.
The contour $C$ is transformed into the contour $C_{\epsilon}$, as shown in
\fig{contour_deformation}, avoiding the branch point at the
origin.  By considering the integration over the contour $C_{\epsilon}$, and
by performing the change of variables $z\to -\mi(z/\delta -\epsilon)$, we have:
\begin{equation}
   \sum_{n=1}^{\infty}
   \frac{\tilde{f}(n\delta)}{n^{\alpha}}=-\frac{\mi}{2}\int_{-\infty}^{\infty}\dd{z}
   \frac{\tilde{f}[\delta(\mi z+\epsilon)]}{(\mi z+\epsilon)^{\alpha}}
   \cothc{\mpi(z-\mi\epsilon)}.
   \elabel{int_su_Ce}
\end{equation}
\begin{figure}
   \centerline{
   \epsfxsize=10.0cm
   \epsfclipon
   \epsfbox{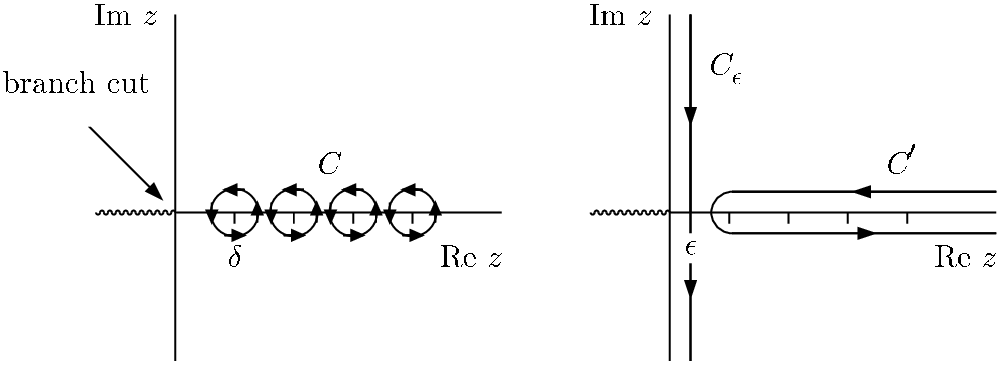}}
   \caption{Contours for the integral in \eq{int_su_C}. The indicated deformation of the
     contour ($C \rightarrow C' \rightarrow C_{\epsilon}$) transforms  \eq{int_su_C} into \eq{int_su_Ce}.}
   \flabel{contour_deformation}
\end{figure}
To extract the $\delta$-series of the above integral, it is tempting
to integrate term by term the Taylor series of the function
$\tilde{f}(z)$. We first split the above integral into the following terms:
\begin{multline*}
\sum_{n=1}^{\infty}
\frac{\tilde{f}(n\delta)}{n^{\alpha}}=-\frac{\mi}{2}\int_{0}^{\infty}\dd{z}
\tilde{f}[\delta(\mi z+\epsilon)](\mi z+\epsilon)^{-\alpha}
\glc \cothc{\mpi(z-\mi\epsilon)}-1 \grc - \frac{\mi}{2}\int_{0}^{\infty}\dd{z}
\tilde{f}[\delta(\mi z+\epsilon)](\mi z+\epsilon)^{-\alpha} \\
-\frac{\mi}{2}\int_{-\infty}^{0}\dd{z}
\tilde{f}[\delta(\mi z+\epsilon)](\mi z+\epsilon)^{-\alpha}
\glc \cothc{\mpi(z-\mi\epsilon)}+1  \grc  +\frac{\mi}{2}\int_{-\infty}^{0} \dd{z}
\tilde{f}[\delta(\mi z+\epsilon)](\mi z+\epsilon)^{-\alpha}.
\end{multline*}
This separation is valid if all the above integrals are well-defined,
a condition respected by the functions $f(z)$ considered in this paper
(for $\delta \le 1$). Using the following representation of the Riemann
Zeta function, valid for $Re(\beta)<1$:
\begin{equation*}
   \zeta(\beta)=\lim_{\epsilon\to0}\glb-\frac{\mi}{2} \glc \int_{0}^{\infty}
   \dd{z}(\mi z+\epsilon)^{-\beta}(\cothc{\mpi(z - \mi\epsilon)}-1)+\int_{-\infty}^{0}
   \dd{z}(\mi z+\epsilon)^{-\beta}(\cothc{\mpi(z-\mi\epsilon)}+1) \grc \grb,
\end{equation*}
we obtain
\begin{equation}
   \sum_{n=1}^{\infty} \frac{\tilde{f}(n\delta)}{n^{\alpha}}\
   =\sum_{m=[\alpha]}^{\infty}\frac{f^{(m)}(0)}{m!}\delta^m \zeta(\alpha-m)+
   \frac{-\mi}{2}\int_{0}^{\infty}\dd{z}
   \tilde{f}(z)(\mi z+\epsilon)^{-\alpha}+\frac{\mi}{2}\int_{-\infty}^{0} \dd{z}
   \tilde{f}(z)(\mi z+\epsilon)^{-\alpha}.
   \elabel{lastcontour}
\end{equation}
Finally, by considering the contours shown in \fig{cont_integ}, we
verify that
\begin{equation*}
   \lim_{\epsilon \to 0} -\frac{\mi}{2}\int_{0}^{\infty} \dd{z}
   \tilde{f}(\delta(\mi z+\epsilon))(\mi z+\epsilon)^{-\alpha}+\frac{i}{2}
   \int_{-\infty}^{0} \dd{z} \tilde{f}(\delta(\mi z+\epsilon))(\mi z+\epsilon)^{-\alpha}=
   \delta^{\alpha-1}\int_{0}^{\infty} \dd{t} \frac{\tilde{f}(t)}{t^\alpha}.
\end{equation*}
\begin{figure}
   \centerline{
   \epsfxsize=3.3cm
   \epsfclipon
   \epsfbox{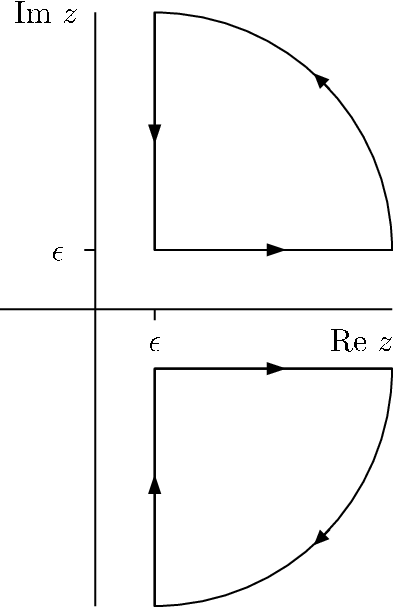}}
   \caption{Contours of the integrals in \eq{lastcontour}.}
   \flabel{cont_integ}
\end{figure}
Collecting these results proves \eq{sum_integral_zeta}.

\subsection{Expansion formula for integer $\alpha$}

The series in \eq{sum_integral_zeta} is not defined for integer $\alpha= m+1$, because of the
simple pole of the Riemann Zeta function $\zeta(z)$  at $z=1$. 
However, this divergence is compensated by an ultraviolet divergence in
the integral. We analyze this situation by considering
$\alpha=\lfloor\alpha\rfloor+\rho$, and by taking the limit
$\rho\to 0$. We write:
\begin{equation}
   \delta^{\lfloor\alpha\rfloor-1 +\rho}\int_{0}^{\infty} \dd{t}
   \glb\frac{f(t)}{t^\alpha} - \sum_{m=0}^{\lfloor\alpha\rfloor-1} \frac{f^m(0)
   t^{m-\lfloor\alpha\rfloor+\rho}}{m!} \grb + \delta^{\lfloor\alpha\rfloor-1}
   \frac{f^{(\alpha-1)}(0)}{(\alpha-1)!}  \zeta(1+\rho).
   \elabel{split_equation}
\end{equation}
In order to isolate the ultraviolet divergence, we split the integral into
two infrared-divergent terms, and \eq{split_equation} becomes 
\begin{equation*}
   \delta^{\lfloor\alpha\rfloor+\rho-1}\int_{\epsilon}^{\infty} \dd{t}
   \glb\frac{f(t)}{t^\alpha} - \sum_{m=0}^{\lfloor\alpha\rfloor-2}
   \frac{f^{(m)}(0) t^{m-\alpha}}{m!} \grb-
   \delta^{\lfloor\alpha\rfloor+\rho-1}\int_{\epsilon}^{\infty} \dd{t}
   \frac{f^{(\alpha-1)}(0)}{(\alpha-1)! t^{1+\rho}} +
   \delta^{\lfloor\alpha\rfloor-1} \frac{f^{(\alpha-1)}(0)}{(\alpha-1)!}
   \zeta(1+\rho),
\end{equation*}
where $\epsilon$ is an infrared cut-off.  Using the expansion
$\zeta(1+\rho)=1/\rho+\gamma+O(\rho)$, where $\gamma$ is the Euler constant,  
we can take the limit $\rho\to 0$:
\begin{equation*}
   \lim_{\rho \to 0} \glc  \zeta(1+\rho) - \delta^{\rho}\int_{\epsilon}^{\infty}
   \dd{t} \frac{1}{t^{1+\rho}} \grc = \glc  \gamma -\log(\delta)
   -\log(\epsilon)  \grc.
\end{equation*}
From the above we get the following expansion formula for integer values of $\alpha$:
\begin{equation}
   \sum_{n=1}^{\infty} \frac{f(n\delta)}{n^{\alpha}} = \delta^{\alpha-1}
   \glb\const - \frac{f^{(\alpha-1)}(0)}{(\alpha-1)!}
   \log(\delta)\grb +\sum_{m\ne \alpha-1} \delta^m  f^{(m)}(0)
   \frac{\zeta(\alpha-m)}{m!}, 
   \elabel{formula_integer}
\end{equation}
where the constant is expressed in term of the following limit:
\begin{equation*}
   \const=\lim_{\epsilon \to 0}\glb\int_{\epsilon}^{\infty} \dd{t}
   \glb\frac{f(t)}{t^\alpha} - \sum_{m=0}^{\alpha-2} \frac{f^{(m)}(0)
   t^{m-\alpha}}{m!} \grb +
   \frac{f^{(\alpha-1)}(0)}{(\alpha-1)!}\glb\log(\epsilon)+
   \gamma\grb\grb.
\end{equation*}

\section{General cumulants} 
\slabel{generic_cumulant}
For completeness, we derive in this appendix the trace formula for the cumulants $\kappa_l$
of the rescaled width distribution $\Psi(s)$ for all $l$ from their generating function
\begin{equation*}
   \Psi(s)=\log \mean{\expb{s w_2}}=\sum_{l=1}^{\infty}\frac{\kappa_{l}}{l!}s^l.
\end{equation*}
We first consider the full periodic series.
To compute the average $\mean{\expb{s w_2}}$ we have to integrate over the
Gaussians $\SET{a_n,b_n}$. To make all the Gaussian integrals 
well-defined, we introduce a cut-off $N$ in the momentum space, such that
$n=1\TO N$. The average $ \mean{\expb{s w_2}}$ takes the form:
\begin{equation*}
   \mean{\expb{s w_2^{\period}}}=\frac{\glb\prod_{n'} \int_{-\infty}^{\infty} 
   \dd{x_{n'}}\grb \expc{-\half\sum_{n'm'} M_{n'm'}(s) x_{n'}
   x_{m'}}}{\glb\prod_{n'} \int_{-\infty}^{\infty} \dd{x_{n'}}\grb
   \expc{-\half \sum_{n'm'} M_{n'm'}(0) x_{n'} x_{m'}}}= \glc \frac{\text{Det}
   M(s)}{\text{Det} M(0)}\grc ^{-\frac{1}{2}},
\end{equation*}
where the indices $n'$ and $m'$ run from $0$ to $2N$. In the above
expression, we set $x_{n'}=a_n'$, for $n'=0 \TO N$ and $x_{n'}=b_{n'-N}$ for
$n'=N+1 \TO 2N$. The matrix $M(s)$ is a block of four
$N \times N$ matrices defined as
\begin{equation*}
   M(s) =\glb \begin{array}{cc}
   \sigma_{n}^{-2}\delta_{nm} & 0 \\ 0 & \sigma_{n}^{-2}\delta_{nm} \end{array}
   \grb+s\glb \begin{array}{cc} 2C_{nm} & I_{nm} \\ I_{mn} & 2 S_{nm}
   \end{array} \grb
\end{equation*}
(for the coefficients $C_{nm}$, $I_{nm}$,  and
$S_{nm}$, see \eq{w2_expr}).  We arrive at the
following expression for the $\Psi(s)$:
\begin{equation*}
   \Psi(s)=-\frac{1}{2}\log \text{Det}
   M(s)+\frac{1}{2}\log \text{Det}
   M(0)=-\frac{1}{2}\Tr \glc \log M(s) \grc +\frac{1}{2}\Tr \glc \log M(0) \grc.
\end{equation*}
Expanding the logarithm $\log M(s) $ 
one obtains
\begin{equation*}
   \Psi(s/\mean{w_2})=\sum_{l=1}^{\infty}
   \frac{(s/\mean{w_2})^{l}}{l}\Tr\glc (A^{\period})^{l} \grc,
\end{equation*}
where $A^{\period}$ is given in \eq{A_per}. 
Hence the $l$th cumulant of the roughness is given by
\begin{equation}
   \kappa^{\period}_{l}=\frac{(l-1)!}{2\mean{w_2}^{l}}\Tr 
   \glc (A^{\period})^{l} \grc.
   \elabel{cumulant_expr_period}
\end{equation}
For the cosine series, one has to integrate over the
Gaussian variables $\SET{c_n}$ and $\SET{s_n}$.
One arrives at an analogous expression for the cumulants.

\end{document}